\documentclass[twocolumn]{jpsj2} 

\title{Multiband Superconductivity in Heavy Fermion Compound CePt$_3$Si without Inversion Symmetry: An NMR Study on a High-Quality Single Crystal}

\author{Hidekazu Mukuda$^{1}$\thanks{E-mail address: mukuda@mp.es.osaka-u.ac.jp}, Sachihiro Nishide$^{1}$, Atsushi Harada$^{1}$, Kaori Iwasaki$^{1}$, Mamoru Yogi$^{1,3}$, Mitsuharu Yashima$^{1}$, Yoshio Kitaoka$^{1}$, Masahiko Tsujino$^{2}$, Tetsuya Takeuchi$^{2}$, Rikio Settai$^{2}$, Yoshichika \={O}nuki$^{2}$,\\ Ernst Bauer$^{4}$, Kohei M. Itoh$^{5}$, E. E. Haller$^{6}$}

\inst{$^{1}$Graduate School of Engineering Science, Osaka University, Toyonaka, Osaka 560-8531 \\
$^2$Department of Physics, Graduate School of Science, Osaka University, Osaka 560-0043\\
$^3$Department of Physics, Faculty of Science, University of the Ryukyus, Nishihara, Okinawa 903-0213\\
$^4$Institute of Solid State Physics, Vienna University of Technology, A-1040 Wien, Austria\\
$^5$Department of Applied Physics and Physico-Informatics, Keio University, Yokohama 223-8522\\
$^6$University of California at Berkeley and Lawrence Berkeley National Laboratory, Berkeley, CA 94720, USA
}

\abst{
We report on novel superconducting characteristics of the heavy fermion (HF) superconductor CePt$_3$Si without inversion symmetry through $^{195}$Pt-NMR study on a single crystal with $T_c=$ 0.46 K that is lower than $T_c\sim$ 0.75 K for polycrystals. 
We show that the intrinsic superconducting characteristics inherent to CePt$_3$Si can be understood in terms of the unconventional strong-coupling state with a line-node gap below $T_c=$ 0.46 K. 
The mystery about the sample dependence of $T_c$ is explained by the fact that more or less polycrystals and single crystals inevitably contain some disordered domains, which exhibit a conventional BCS $s$-wave superconductivity (SC) below 0.8 K.
In contrast, the N\'eel temperature $T_{\rm N}\sim 2.2$ K is present regardless of the quality of samples, revealing that the Fermi surface responsible for SC differ from that for the antiferromagnetic order. 
These unusual characteristics of CePt$_3$Si can be also described by a multiband model; in the homogeneous domains, the coherent HF bands are responsible for the unconventional SC, whereas in the disordered domains the conduction bands existing commonly in LaPt$_3$Si may be responsible for the conventional $s$-wave SC. 
We remark that some impurity scatterings in the disordered domains break up the 4f-electrons-derived coherent bands but not others. 
In this context, the small peak in $1/T_1$ just below $T_c$ reported in the previous paper (Yogi et al,  (2004)) is not due to a two-component order parameter composed of spin-singlet and spin-triplet Cooper pairing states, but due to the contamination of the disorder domains which are in the $s$-wave SC state.
}

\kword{non-centrosymmetric superconductivity, CePt$_3$Si, antiferromagnetism, NMR}

\begin{document}
\maketitle

\date{\today}


\section{Introduction}

For a superconducting order parameter, it is generally believed that time reversal invariance is a necessary condition for the formation of spin-singlet Cooper pairing, while an inversion center is additionally required for the formation of spin-triplet pairing \cite{Anderson}. In superconductors that lack inversion symmetry, the relationship between spatial symmetry and the Cooper-pair spin state may be broken, which will make the parity of superconducting state mixed between even and odd parities due to the  antisymmetric spin-orbit coupling (ASOC) \cite{Gorkov,Frigeri,FrigeriNJ,Samokhin,Hayashi,Fujimoto,Yanase}, thereby leading to a two-component order parameter composed of spin-singlet and spin-triplet Cooper pairing states. 
The first heavy fermion (HF) superconductivity (SC) without inversion symmetry was discovered in CePt$_3$Si\cite{Bauer}.
After that, the pressure-induced superconductivity of this family was also observed for a ferromagnet UIr and antiferromagnets CeRhSi$_3$\cite{Kimura}, CeIrSi$_3$\cite{Sugitani}, and CeCoGe$_3$\cite{Settai}. 
An intriguing feature of these systems is that the splitting of the Fermi surfaces by ASOC is much larger than the superconducting gap\cite{Samokhin,Hashimoto}.
Interestingly, in the SC state of CeIrSi$_3$ and CeRhSi$_3$, it has been reported that the upper critical fields ($H_{c2}$'s) exceed 30 T along the c-axis, which are extremely high as compared to a Pauli limiting field.\cite{KimuraPRL2007,Okuda,SettaiHc2} 
This kind of behavior has never been observed in the other HF superconductors such as CeCoIn$_5$, CeIrIn$_5$, and CeCu$_2$Si$_2$, which are in a spin-singlet Cooper pairing regime. 
In the noncentrosymmetric superconductors where the correlation between electrons may be not so significant, however, the characteristic feature for conventional $s$-wave spin singlet superconducting state has been observed in many compounds, for example, Y$_2$C$_3$\cite{Harada,Akutagawa}, Ir$_2$Ga$_9$\cite{Takagi,Harada29}, LaPt$_3$Si\cite{Yogi2_2006}, LaIrSi$_3$\cite{MukudaCe113}, Li$_2$Pd$_3$B\cite{Nishiyama1}, and so on, whereas the parity mixing state has been argued only in the related compound Li$_2$Pt$_3$B\cite{Yuan,Nishiyama}.

The CePt$_3$Si exhibits superconductivity at $T_{\rm c}=$ 0.75 K in an antiferromagnetically ordered state below a N\'{e}el temperature of $T_{\rm N}=$ 2.2 K, as reported by Bauer {\it et al.} \cite{Bauer,BauerReview} 
Neutron-scattering measurement probed an AFM structure with a wave vector $Q= (0,0,1/2)$ and a magnetic moment of 0.16$\mu_B$ lying in the {\it ab}-plane of the tetragonal lattice \cite{Metoki}.
Uniform coexistence of the AFM order and SC has been microscopically evidenced by NMR \cite{Yogi2004,Yogi2006} and $\mu$SR \cite{Amato}. 
Although CePt$_3$Si has attracted considerable attention as the first superconductor without inversion symmetry \cite{Frigeri,FrigeriNJ,Samokhin,Hayashi,Fujimoto,Yanase}, the experimental results are still contradictory. 
For example, the London penetration depth \cite{Bonalde}, thermal conductivity \cite{Izawa}, and NMR \cite{Yogi2004} revealed the line node in the SC gap function well below $T_{\rm c}$, whereas the small peak that was observed in $1/T_1$ just below $T_{\rm c}$ provided evidence for the inclusion of an $s$-wave component without nodes \cite{Yogi2004}. 
However, the evidence of parity mixing due to the lack of inversion symmetry has not been obtained in this compound so far, suggesting that the theoretical models for CePt$_3$Si put forth to date \cite{Hayashi} seem to be inadequate because of the unknown complicated multiband effect. 
The $T_c$ value, however, still remains a mystery \cite{Scheidt,Kim,Nakatsuji,Motoyama,Aoki}; Takeuchi and coworkers reported that for a high-quality single crystal of CePt$_3$Si, specific-heat measurements revealed a bulk SC at $T_c=$ 0.46 K \cite{Takeuchi}, which is remarkably lower than $T_c=$ 0.75 K for a polycrystal \cite{Bauer}.  Despite the lower $T_c$ of the sample, its quality is guaranteed by the extremely small value of a residual $\gamma$ term, i.e., the $T$-linear coefficient of the electronic specific heat well below $T_c$ is smaller than that for polycrystals \cite{Takeuchi}.  In this high-quality single crystal, however, the resistivity drops to zero below 0.75 K, similar to the case of polycrystals. There still remains an underlying mystery, namely, the sample dependence of $T_c$ prevents us from identifying any intrinsic SC properties of CePt$_3$Si. 
In particular, it is also crucial to reveal the origin of the small peak in $1/T_1$ just below $T_{\rm c}$ in the previous sample with $T_{\rm c} \sim 0.75$ K \cite{Yogi2004}. 

On the basis of $^{195}$Pt-NMR studies on a high-quality single crystal, we report in this paper that the genuine SC characteristics inherent to CePt$_3$Si can be understood in terms of an unconventional strong-coupling state with a line-node gap below $T_c=$ 0.46 K; further, the mystery about the sample dependence of $T_c$ is explained by the fact that the disordered domains with $T_c=$ 0.8 K are inevitably contained even in a single crystal. Then, we propose a possible multiband model to account for the disparate SC characteristics and the universal presence of the AFM order in CePt$_3$Si. This model allows us to consider that SC and the AFM order occur on different pieces of Fermi surfaces, but the 4f-electrons-derived coherent HF bands collapse partially due to the disorders inevitably included in the samples. 
In this context, the small peak in $1/T_1$ just below $T_c$ reported in the previous paper \cite{Yogi2004} is not due to a two-component order parameter composed of spin-singlet and spin-triplet Cooper pairing states, but due to the contamination of the disorder domains that are in the s-wave SC state.  
Furthermore, we compare the results on CePt$_3$Si with that of another noncentrosymmetric HF compound CeIrSi$_3$. 
As a result, we remark that multiband and single band pictures may be applicable for understanding the phase diagrams of AFM and SC in CePt$_3$Si and CeIrSi$_3$, respectively, leading to an indication that the higher SC transition tends to emerge for a single band rather than for multibands among the various HF compounds.


\section{Experimental}

A high-quality single crystal ($\sharp4$) of CePt$_3$Si was grown by the Bridgman method in a Mo crucible, as described elsewhere \cite{Takeuchi}. A very small piece of the crystal aligned along the (001) tetragonal plane was cut out from the ingot of a large single crystal by using the x-ray Raue method. The quality of this sample was guaranteed by a sharp jump in the specific heat at $T_c=$ 0.46 K, and by the small value of the residual $\gamma$ term, which was less than 34 mJ/K$^{2}$mol well below $T_c$. The resistivity of the sample began to decrease below 0.8 K; at 0.75 K, the resistivity became zero, as reported in literature \cite{Takeuchi}. For NMR measurements, this crystal ($\sharp4$) was moderately crushed into coarse powder; the particles were approximately more than a few hundred micrometers in diameter and were oriented with the magnetic field parallel to the c-axis. 

We examined the sample dependence of SC and the magnetic properties of CePt$_3$Si via NMR studies. 
The samples used in the NMR studies are listed in Table I along with their $T_c$ values estimated from the jump in the specific heat and the drop in the resistivity. The previous NMR investigations by Yogi {\it et al.}\cite{Yogi2004,Yogi2006} were carried out using a single crystal ($\sharp$2), as in literature \cite{Yasuda}. The specific-heat measurement of this sample revealed a broad increase below 0.75 K, and a small peak at 0.46 K (see Fig. 4); further, the resistivity dropped to zero below 0.75 K. 
It should be noted that the polycrystal sample ($\sharp$1) used in the first report by Bauer {\it et al.}\cite{Bauer} exhibited a broad peak in the specific heat and zero resistance around $T_c=$ 0.75 K, but did not reveal any anomalies around 0.46 K.
In order to examine the possible disorder effects arising from lattice distortion, the NMR study was also carried out by crushing coarse powder of $^{29}$Si-enriched polycrystal ($\sharp$3a) into well-ground powder ($\sharp$3b).  
The average diameters of $\sharp$3a and $\sharp$3b were approximately a few hundred micrometers and less than $20\mu$m, respectively. 
In particular, the sample $\sharp$3b is anticipated to include many local lattice distortions or crystal defects in the shorter length scale than the diameter of the particles due to the thoroughly crushing process over several hours.  
It should be noted that the temperature ($T$) dependence of the specific heat of the samples ($\sharp$3a) and ($\sharp$3b) are different, as will be shown in Fig. \ref{fig:specificheat}. Thus, unexpected sample dependence of the SC property is one of the main issues in this study, as discussed later.

\section{Results and Analyses}

\subsection{$^{195}$Pt-NMR Spectra }

\begin{figure}[htbp]
\begin{center}
\includegraphics[width=0.6\linewidth]{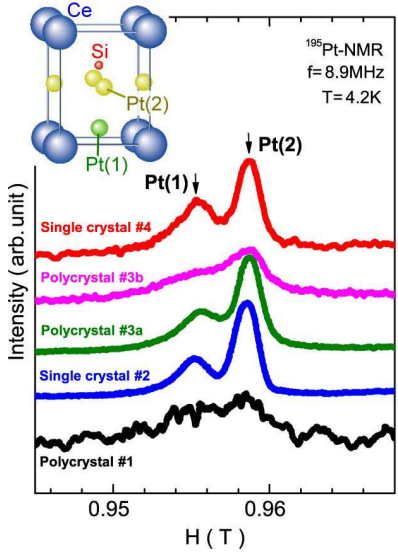}
\end{center}
\caption[]{(Color online) $^{195}$Pt-NMR spectra at 8.9 MHz and 4.2 K for the single crystal $\sharp$4 and other crystals. The inset indicates the atomic configuration in a unit cell of CePt$_3$Si without inversion symmetry.  Two inequivalent crystallographic Pt sites in a unit cell $-$ denoted by Pt(1) and Pt(2) $-$ are distinguished by the difference in their Knight shifts.\cite{Yogi2004} The NMR spectral widths for the single crystal $\sharp$4, the single crystal $\sharp$2, and the polycrystal $\sharp$3a are narrower than those for the polycrystal $\sharp$1 \cite{Bauer} and the well-ground powder sample $\sharp$3b, indicating that the samples $\sharp$1 and $\sharp$3b include some kind of disorders.}
\label{fig:spectra}
\end{figure}

In order to characterize the samples used for the NMR studies, we present the field-swept $^{195}$Pt-NMR spectra for the single crystal ($\sharp4$) and other crystals in Fig.\ref{fig:spectra}. Two inequivalent crystallographic Pt sites denoted by Pt(1) and Pt(2) \cite{Yogi2004} are distinguished by the difference in their Knight shifts. The NMR spectral widths for the single crystals $\sharp$4 and $\sharp$2 and the polycrystal $\sharp$3a are narrower than those for the polycrystal $\sharp$1 and the well-ground powder sample $\sharp$3b; from these spectral widths we can observe that the single crystal $\sharp$4 has better quality on a microscopic level. 
In contrast, polycrystals $\sharp$1 and $\sharp$3b are affected by some disorders such as impurities, crystal defects, local-lattice distortions, and so on. It should be noted that 
the $T$ dependence of the $^{195}$Pt Knight shift for the present single crystal $\sharp4$ coincides with that of $\sharp2$ \cite{Yogi2004,Yogi2006}. 

\subsection{Nuclear spin-lattice relaxation rate}

\begin{figure}[htbp]
\begin{center}
\includegraphics[width=0.8\linewidth]{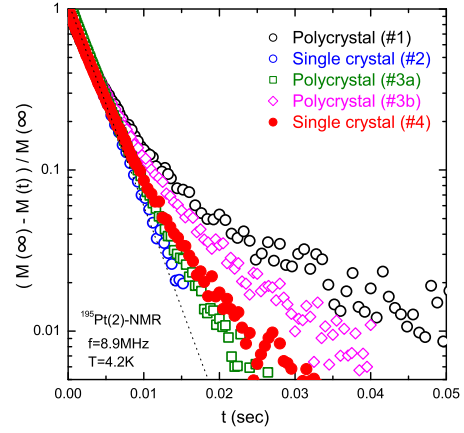}
\end{center}
\caption[]{(Color online) Recovery curves of nuclear magnetization $m(t)$ (see text) at the Pt(2) site for the single crystal $\sharp$4 and other samples at 4.2 K. It should be noted that all $m(t)$ curves cannot be fitted by a single exponential function (dotted line), but by a multiexponential function. A similar behavior was also observed at the Pt(1) and Si sites, suggesting that the electronic states of all the samples were more or less distributed over each sample, depending on a fraction of the disordered domains into which defects were inevitably or intentionally introduced. The long component ($T_{1L}$) becomes more prominent for the polycrystal $\sharp$1 and the well-ground powder sample $\sharp$3b, which represents the large volume fractions of the disordered domains in their crystals.}
\label{fig:recovery}
\end{figure}

The measurements of the nuclear spin-lattice relaxation rate $1/T_1$ allow us to characterize each sample. When an electronic state is homogeneous over a sample, the recovery curve of $^{195}$Pt nuclear magnetization ($I=1/2$) is generally determined by a simple exponential function given by, 
\[
m(t)\equiv\frac{M(\infty)-M(t)}{M(\infty)}=\exp\left(-\frac{t}{T_{1}}\right),
\]
where $M(\infty)$ and $M(t)$ are the nuclear magnetization at the thermal equilibrium condition and the nuclear magnetization at a time $t$ after the saturation pulse, respectively. However, as indicated in Fig. \ref{fig:recovery}, $m(t)$ cannot be fitted by a single exponential function for all the crystals. 
Unexpectedly, even in the high-quality sample $\sharp$4, the $m(t)$ cannot be fitted by a single exponential function; however, it can be fitted by a multiexponential function over the entire temperature range. 
Such behavior was also observed in the recovery curves at both the Pt(1) and Si sites, indicating the presence of some inhomogeneity in the electronic states over the sample, which are spatially distributed  over a macroscopic scale.  It is noteworthy that the fraction of long components in $T_1$ is larger for the polycrystal $\sharp$1 than for the single crystal $\sharp$4. 
Furthermore, the fraction of the long components of the well-ground powder sample $\sharp$3b becomes larger than that for the coarse powder $\sharp$3a. These facts reveal that the longer components arise from the domains affected by some disorders such as impurities,  crystal defects, local-lattice distortions, and so on. 

The recovery curve of the single crystal $\sharp$4 is tentatively fitted by assuming two components in $T_{1}$, i.e., short and long components denoted by $T_{1S}$ and $T_{1L}$, respectively, as follows: 
\[
m(t)=A_S\exp\left(-\frac{t}{T_{1S}}\right)+A_L\exp\left(-\frac{t}{T_{1L}}\right),
\]
where $A_S$ and $A_L$ are the respective fractions of the short and long components, and  $A_S+A_L=1$. 
It should be noted that $A_S\approx$ 0.7 and $A_L\approx$ 0.3 for the single crystal $\sharp$4  correspond to the respective volume fractions of the homogeneous and disordered domains in the sample; hence, it can be stated that these fractions depend on the samples. 

\begin{figure}[htbp]
\begin{center}
\includegraphics[width=0.8\linewidth]{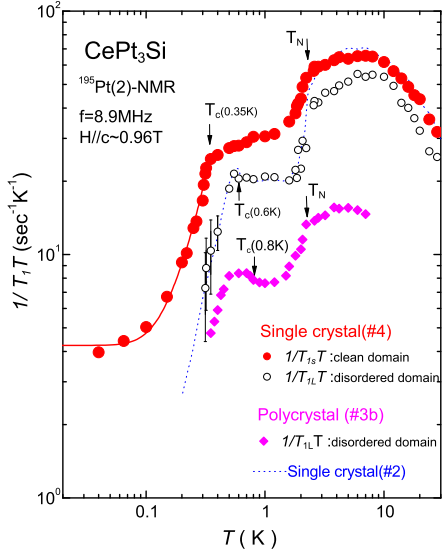}
\end{center}
\caption[]{(Color online) $T$ dependence of $1/T_{1S}T$ and $1/T_{1L}T$ at the Pt(2) site of the single crystal $\sharp4$. $1/T_{1S}T$ derived from the homogeneous domains evidences the unconventional SC state with the line-node gap below $T_{\rm c}(H)=$ 0.35 K at $H_c$=0.96 T along the c-axis. In contrast, $1/T_{1L}T$ derived from the disordered domains decreases with a tiny peak just below $T_c(H)\sim$ 0.6 K, which is the same as the $T_c$ for the polycrystal at $H\sim0.96$ T, similar to the previous result for the single crystal  $\sharp$2 (dotted line) obtained by Yogi {\it et al.}\cite{Yogi2004}}
\label{fig:T1}
\end{figure}

\subsection{Intrinsic SC properties inherent to CePt$_3$Si \\ $-$ short $T_1$ component of single crystal $\sharp4$ $-$ }  

First we note a short component $T_{1S}$ that is more dominant for the high-quality single crystal($\sharp4$), which represents an intrinsic property inherent to CePt$_3$Si. 
Figure \ref{fig:T1} shows the $T$ dependence of $1/T_{1S}T$ at $H\sim$ 0.96 T along the c-axis. 
It decreases below $T_{\rm N}= 2.2$ K and drops steeply below $T_c(H)\approx$ 0.35 K  at $H\sim$ 0.96 T for the single crystal with $T_{\rm c}=$ 0.46 K at $H=0$ \cite{Takeuchi}. Interestingly, in the SC state, the $1/T_{1S}$ shows a $T^3$-like dependence without a coherence peak just below $T_{\rm c}$, followed by a $T_1T=const$-like behavior well below $T_{\rm c}$. 
These facts are evidence of the unconventional superconducting nature of the intrinsic domain of CePt$_3$Si. 
By assuming a line-node gap model with $\Delta=\Delta_0\cos\theta$ and the residual density of states (RDOS) at the Fermi level $N_{\rm res}(E_{\rm F})$,  $1/T_{1S}T$ in the SC state is well reproduced by
\[
\frac{1/T_{1S}T}{1/T_{1S}T_{\rm c}}=\frac{2}{k_{\rm B}T}\int\left(\frac{N_{\rm s}(E)}{N_0}\right)^2 f(E)[1-f(E)]dE,
\]
where $N_{\rm s}(E)/N_0=E/\sqrt{E^2-\Delta^2}$; $N_0$ is the DOS at $E_{\rm F}$ in the normal state and $f(E)$ is the Fermi distribution function. As shown by the solid line in Fig. \ref{fig:T1}, the experimental result is well fitted by assuming $2\Delta_0/k_{\rm B}T_{\rm c}\approx$ 6 and $N_{\rm res}/N_0\approx$ 0.41, suggesting that strong-coupling SC emerges with line-node gap. 
It is expected that the large RDOS in the SC state is caused due to the impurity effect and/or the Volovik effect where the RDOS is induced by a supercurrent in the vortex state \cite{Volovik}.  
Since the contribution of the former is estimated to be less than 10\% of the DOS at $T_c$ from specific-heat measurement \cite{Takeuchi}, the contribution of the latter may be dominant in the present case.

\subsection{Disordered domain in the single crystal \\ $-$ long $T_1$ component of single crystal $\sharp4$ $-$}  

\begin{figure}[htbp]
\begin{center}
\includegraphics[width=0.8\linewidth]{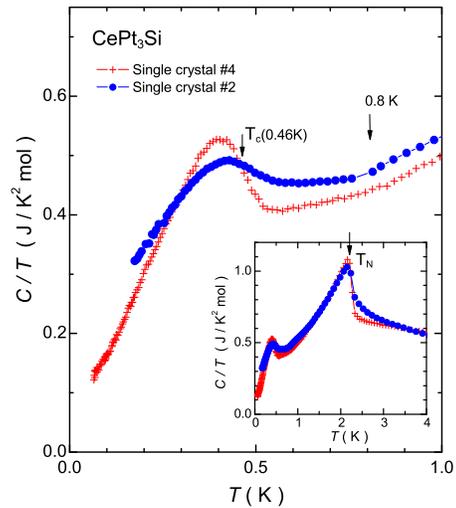}
\end{center}
\caption[]{(Color online) The specific heat in the single crystal $\sharp2$ used in the previous NMR study \cite{Yogi2004} along with that in the present crystal $\sharp4$ [cited from ref.\cite{Takeuchi}]. It exhibits the gradual increase of $C/T$ between 0.4 K and 0.8 K upon cooling and broader peak at 0.46 K than in the present single crystal $\sharp4$, suggesting that the sample $\sharp$2 actually includes large inhomogeneity in the SC property due to the presence of the disordered domains. }
\label{fig:SH_singles}
\end{figure}

The $T$ dependence of the long component $1/T_{1L}T$ for the single crystal $\sharp4$ is different from that of the short $T_{1}$ component. 
As shown in Fig. \ref{fig:T1}, the $1/T_{1L}T$ decreases below $T_{\rm N}= 2.2$ K, and there is a marked decrease in  $1/T_{1L}T$ with a tiny peak just below $T_c(H)\sim$ 0.6 K that is the value in the field of $H\sim0.96$ T for the polycrystal  with $T_c(H=0)= 0.75$ K \cite{Bauer}. 
This $T$ variation of $1/T_{1L}T$ resembles the result reported previously for the single crystal $\sharp$2 \cite{Yogi2004}, as shown by the dotted line in Fig. \ref{fig:T1}. 
Although we have claimed that the $1/T_1$ in the single crystal $\sharp$2 has a peak just below $T_c$ in the previous paper\cite{Yogi2004}, we should note that the previous result represents the $1/T_1$ data of the disordered domains accidentally contained. 
In fact, the previous single crystal $\sharp$2 includes large inhomogeneity in the SC property, which was suggested by the specific heat measurement that exhibits the gradual increase of $C/T$ between 0.4 K and 0.8 K upon cooling and broader peak at 0.46 K than in the present single crystal $\sharp4$ \cite{Takeuchi}, as shown in Fig. \ref{fig:SH_singles}.
If we extract $1/T_{1S}T$ in the sample $\sharp$2 by assuming two $T_1$ components, its $T$-dependence  resembles the present results on the sample $\sharp$4. 
Eventually, as will be discussed about the well-ground powder sample $\sharp3b$ in the next section, the origin of the small peak of $1/T_1$ for the single crystal $\sharp2$ \cite{Yogi2004} is relevant with the coherence peak just below $T_c$($\sim$0.8 K) in the disorder-rich domains. 
Nevertheless, it is puzzling that $T_c=0.75$ K for the disordered domains is higher than that for the homogeneous domains, namely, $T_c=0.46$ K, despite the fact that disorders generally decrease $T_c$ in unconventional HF superconductors. 

\subsection{SC characteristics of disorder-rich domains \\ $-$ long $T_1$ of a well-ground powder sample $\sharp$3b $-$}  

In order to shed light on the SC characteristics of disordered domains, we examined a well-ground powder sample $\sharp$3b into which disorders were intentionally introduced. 
The presence of disorder in this sample was actually suggested by the facts of the significant line broadening of the NMR spectrum (see Fig. \ref{fig:spectra}) and a large fraction of long $T_1$ components  in the recovery curve (see Fig. \ref{fig:recovery}). 
The $m(t)$ of this sample ($\sharp$3b) comprises the multi $T_{1}$ components due to the distribution of the disordered domains. 
A predominant component of the long $T_1$s is extracted from the $m'(t)$ that is obtained by subtracting the contribution of short $T_{1S}$ component, which represents the SC characteristics of the disorder-rich domains.

\begin{figure}[htbp]
\begin{center}
\includegraphics[width=0.8\linewidth]{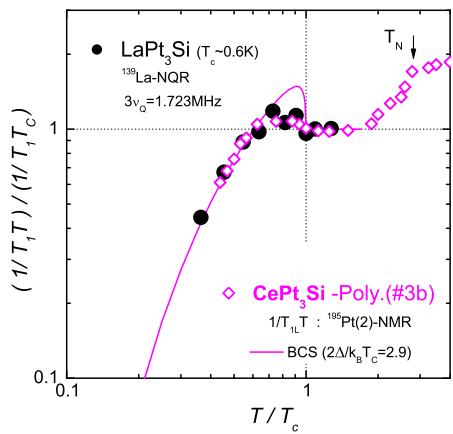}
\end{center}
\caption[]{(Color online) $1/T_{1L}T$ normalized by the value at $T_c$ for sample $\sharp$3b, whose electronic states are dominated by disordered domains.  A distinct coherence peak in $1/T_{1L}T$ just below $T_{\rm c}\approx 0.8$ K and the exponential decrease in $1/T_{1L}$ well below $T_{\rm c}$ reveal that the disordered domains are in the BCS $s$-wave SC regime, resembling the $s$-wave superconductor LaPt$_3$Si with $T_c=$ 0.6 K.\cite{Yogi2_2006} The solid curve indicates the result calculated by means of an isotropic $s$-wave model with 2$\Delta_0/k_BT_c=2.9$. }
\label{fig:LaPt3Si}
\end{figure}

In the normal state, the $1/T_{1L}T$ of the disordered domain of the sample $\sharp$3b decreases below $T_{\rm N}\sim$ 2.2 K and stays a constant just above 0.8 K, as shown in Fig. \ref{fig:T1}. 
Remarkably, a distinct coherence peak appears just below $T_{\rm c}\approx 0.8$ K, followed by an exponential-like decrease in $1/T_{1L}$ well below $T_{\rm c}$. 
Unexpectedly, this result resembles the result of $1/T_1T$ of the isostructural compound LaPt$_3$Si with $T_c=$ 0.6 K. 
Figure \ref{fig:LaPt3Si} shows the $1/T_{1L}T$ normalized by the value at $T_c$ for the disordered domains of well-ground powder $\sharp$3b in comparison with that of LaPt$_3$Si  investigated by $^{135}$La-NQR at 3$\nu_Q$($\pm7/2\Leftrightarrow \pm5/2$)= 1.723 MHz\cite{Yogi2_2006}. As shown by the solid curve in the figure, these results are reproduced by an isotropic gap model with $2\Delta_0/k_{\rm B}T_{\rm c}\approx 2.9$ in a weak coupling regime of a BCS-type $s$-wave superconductor mediated by the electron-phonon interaction. 

\begin{figure}[htbp]
\begin{center}
\includegraphics[width=0.7\linewidth]{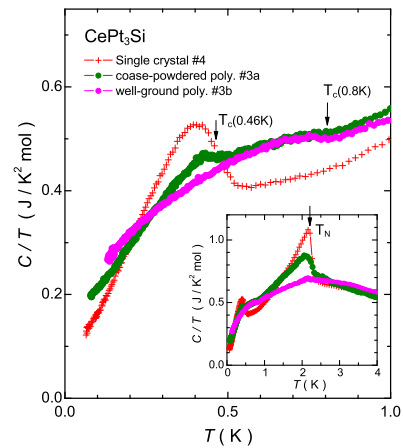}
\end{center}
\caption[]{(Color online)  $T$ dependence of specific heat divided by temperature for the well-ground powder sample $\sharp$3b and the coarse powder sample $\sharp$3a, along with the result for the single crystal $\sharp$4 of CePt$_3$Si. The small jump in $C/T$ at 0.46 K in the coarse powder $\sharp$3a disappears in the well-ground powder sample $\sharp$3b. The $T$ dependence of $C/T$ for sample $\sharp$3b is similar to that for the polycrystal $\sharp$1 first reported by Bauer {\it et al.}.\cite{Bauer} This result is consistent with the $T_1$ result in that the fraction of $T_{1L}$ increases in the case of the sample $\sharp$3b and the polycrystal $\sharp$1, both of which are more affected by disordered domains 
than the others are, as shown in Fig. \ref{fig:recovery}.}
\label{fig:specificheat}
\end{figure}

Here, we note that the values of $T_{\rm c}$ for the disordered domains of CePt$_3$Si, namely, 0.8 K, is equal to the SC onset temperature in resistivity and specific-heat measurements for most CePt$_3$Si crystals \cite{Bauer,Scheidt,Kim,Nakatsuji,Motoyama,Aoki,Takeuchi}. It is suggested that the drop of the resistivity in CePt$_3$Si below 0.8 K takes place in the disordered domains that are inevitably contained. 
It is noteworthy that the disorders introduced by grinding the crystals of CePt$_3$Si into fine powder increase the domains with higher SC transition, namely, $T_c\approx$ 0.8 K. 
Indeed, the increase of bulk $T_c$ is also corroborated by the specific heat measurement for the well-ground powder sample $\sharp$3b. As shown in Fig. \ref{fig:specificheat}, a broad jump in $C/T$ was observed at 0.8 K in the case of sample $\sharp$3b after grinding; the coarse-powdered polycrystal sample $\sharp$3a before grinding exhibited a small jump in $C/T$ at 0.46 K in addition to a gradual increase in $C/T$ below 0.8 K. 
In the case of sample $\sharp$3b, which contains more disordered domains, the residual $\gamma$ term increases well below $T_c$. 
In particular, the $T$ dependence of $C/T$ for the sample $\sharp$3b is quite similar to that for the polycrystal sample $\sharp$1 reported by Bauer {\it et al.} \cite{Bauer} This result is consistent with the results of the NMR spectrum (Fig. 1) and the recovery curve (Fig. 2).  
From these results, the SC state with high $T_c$ (0.8 K) for the disordered domains is consistent with the characteristics of Anderson's dirty superconductor theorem for the $s$-wave Cooper pairing state with an isotropic gap \cite{AndersonDS}. 
The effects of the disorders on thoroughly crushing the LaPt$_3$Si sample is not observed from the recovery curve and the $1/T_1$ data, which suggests that a conduction band inherent to LaPt$_3$Si is not affected by the disorders. 
This is also true for the SC state of the disordered domains of CePt$_3$Si. Nevertheless, it should be noted that the conventional SC state of the disordered domains of CePt$_3$Si coexists with the AFM order on a microscopic scale, as evidenced by the decrease in $1/T_{1L}T$ below $T_{\rm N}\sim$ 2.2 K, as shown in Fig. \ref{fig:T1}. 
In this context, we suppose that the coexistence of SC and the AFM order in the disordered domains in CePt$_3$Si resembles the magnetic superconductors in ternary compounds RERh$_4$B$_4$ (RE = rare earth), where the conventional SC derived from the conduction electron coexists with the localized $4f$-electrons-derived magnetic order \cite{RERh4B4}.  Thus, it should be stated that the SC state of the disordered domains in CePt$_3$Si is not entirely the same as that of LaPt$_3$Si. In fact, the $s$-wave SC in CePt$_3$Si survives under the application of magnetic field of 0.96 T parallel to the c-axis, whereas that of LaPt$_3$Si is suppressed even by $H=$ 0.1 T\cite{Takeuchi}, revealing that the coherence length and the penetration depth of the disordered domains of CePt$_3$Si would be different from those of LaPt$_3$Si. 

\subsection{Pressure effect on CePt$_3$Si: \\ a $^{29}$Si-NMR of $^{29}$Si-enriched polycrystal($\sharp3$a)}

\begin{figure}[htbp]
\begin{center}
\includegraphics[width=0.8\linewidth]{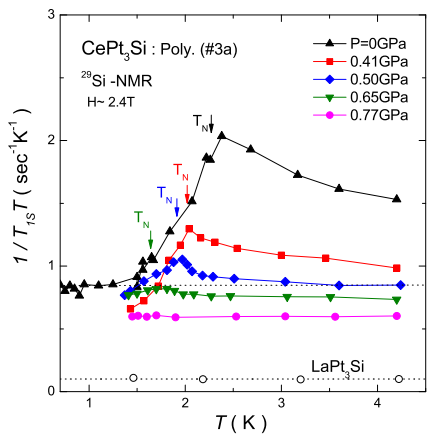}
\end{center}
\caption[]{(Color online) $T$ dependence of $1/T_{1S}T$ measured by $^{29}$Si-NMR for the polycrystal $\sharp$3a of CePt$_3$Si under $P$. The $1/T_{1S}T$ dominated by the AFM spin fluctuations above $T_N$ at $P=0$ is completely suppressed for $P>0.7$ GPa, in association with the drastic decrease of $T_N$ determined by the peak of $1/T_{1S}T$. The DOS at $E_{\rm F}$ below $T_N$, which is proportional to the value of $\sqrt{1/T_{1S}T}$ for $T_{\rm c}<T\ll T_{\rm N}$, does not change significantly with the pressure. 
This result indicates that the collapse of the AFM order does not influence the Fermi surfaces that are responsible for the onset of SC. }
\label{fig:Pressure}
\end{figure}

Next, we address the pressure effect on the AFM order in CePt$_3$Si through $^{29}$Si-NMR measurements of $1/T_{1S}T$ under $P$ for the homogeneous domains of the $^{29}$Si-enriched polycrystal $\sharp$3a. The AFM order of CePt$_3$Si is completely suppressed when $P>0.7$ GPa \cite{Yasuda,Tateiwa,Takeuchi} (see Fig. \ref{fig:phasediagram}(a)), whereas its $T_c$ decreases slightly around an AFM quantum critical point (QCP). As shown in Fig. \ref{fig:Pressure}, the $1/T_{1S}T$ enhanced by the AFM spin fluctuations above $T_N$ is gradually reduced by applying $P$ and is completely suppressed when $P>0.7$ GPa, in association with the drastic decrease in $T_N$ determined by the peak of $1/T_{1S}T$. 
An interesting observation is that the value of $1/T_{1S}T$ for $T_{\rm c}<T\ll T_{\rm N}$ is $\sim$0.8 [sec$^{-1}$K$^{-1}$] when $P=$ 0, which is comparable to $\sim$0.6 [sec$^{-1}$K$^{-1}$] when $P=$ 0.77 GPa, irrespective of the drastic suppression of the AFM order. 
Since the fraction of DOS at $E_{\rm F}$ below $T_N$ is proportional to the value of $\sqrt{1/T_{1S}T}$ for $T_{\rm c}<T\ll T_{\rm N}$, it is revealed that the DOS at the Fermi surface, which is responsible for the onset of SC, does not change when the AFM order collapses. This result reveals that the Fermi surface relevant with the onset of SC differs from that with the AFM order. 

\section{Discussion}

\subsection{Multiband scenario for CePt$_3$Si}

The SC properties of CePt$_3$Si are unusual. The SC inherent to CePt$_3$Si is unconventional with a line-node gap of $T_c=0.46$ K; however, the disorders suppress this intrinsic SC, instead, the $s$-wave SC emerges as a result of the increase of the disordered domains, thereby resulting in a high value of $T_c$ of almost 0.8 K. 
In contrast with the SC characteristics that differ according to the quality of the samples, the AFM order with $T_{\rm N}\sim 2.2$ K is robust against the disorders. 
These results reveal that the Fermi surfaces relevant with the onset of SC differs from that with the AFM order and that the disorders suppress the unconventional SC emerging in the 4f-electrons-derived HF bands inherent to CePt$_3$Si. 
These facts allow us to account for by using a multiband model, that is, there exist three characteristic bands at least. 
The first one is a localized $f$-electron band far below the Fermi level that causes the long-range AFM order below 2.2 K. 
It should be noted that the large reduction in $1/T_1$ points to a possible opening of the gap at the Fermi level below $T_N$. The second one is the $f$-electrons-derived coherent HF bands that are derived from the formation of a periodic lattice of Ce atoms. The third one is a weakly correlated conduction band which leads to the onset of $s$-wave SC in LaPt$_3$Si. 
In the homogeneous domains in samples, the coherent HF bands, which are formed through the hybridization between $f$ electrons and conduction electrons, are responsible for the unconventional SC. Since the disorders break up the coherence of the periodicity of Ce atoms to suppress the unconventional SC inherent to the HF bands, eventually, the conduction bands in the inhomogeneous domains commonly found in LaPt$_3$Si may be responsible for the conventional $s$-wave SC possibly mediated by the electron-phonon interaction. In this model, if the sample were ideally homogeneous being free from any disorders, the conventional BCS $s$-wave SC taking place on the conduction bands may be unfavorable due to the pair-breaking originating from the strong electron correlation effect through an interband coupling with the HF bands. 
Consequently, it is likely that the decrease in resistivity below $T_c\approx$ 0.8 K is due to the inevitably introduced disordered domains. 
Simultaneously, the specific heat shows a broad peak or a double peak between $T_c=$ 0.46 K for the homogeneous domains and $T_c\approx$ 0.8 K for the disordered domains \cite{Bauer,Scheidt,Kim,Nakatsuji,Takeuchi}, which depends on a fraction of the disordered domains to the homogeneous ones. 
This is a possible explanation to settle several underlying issues reported thus far, for instance, the sample dependence of $T_c$.

\subsection{Comparison of the SC characteristics of CePt$_3$Si with that of CeIrSi$_3$}

\begin{figure}[htbp]
\begin{center}
\includegraphics[width=0.8\linewidth]{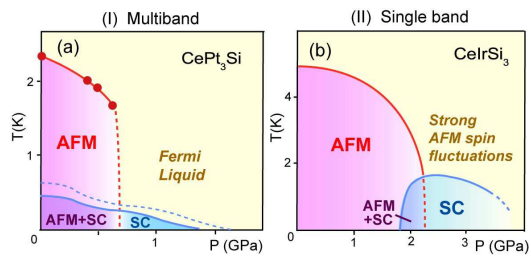}
\end{center}
\caption[]{(Color online) The $P$-$T$ phase diagram for (a) CePt$_3$Si \cite{Yasuda,Tateiwa,Takeuchi} and (b) CeIrSi$_3$\cite{Sugitani}. The SC inherent to CePt$_3$Si emerges under the Fermi-liquid state without any trace of AFM spin fluctuations, whereas that of CeIrSi$_3$ occurs under the Fermi liquid state with strong AFM spin fluctuations.}
\label{fig:phasediagram}
\end{figure}

Here, the SC of CePt$_3$Si is compared with the pressure($P$)-induced SC of CeIrSi$_3$, since both compounds lack inversion symmetry along the c-axis. Interestingly, as illustrated in Figs. \ref{fig:phasediagram}(a) \cite{Yasuda,Tateiwa,Takeuchi} and \ref{fig:phasediagram}(b)\cite{Sugitani}, the $P-T$ phase diagrams of AFM and SC of CePt$_3$Si and CeIrSi$_3$ are very different. 
It should be noted that the $T_c$ for CeIrSi$_3$ is enhanced up to 1.6 K by virtue of the presence of strong AFM spin fluctuations around QCP.\cite{MukudaCe113} 
In the case of CePt$_3$Si, $T_{\rm c}$ ($=$0.46 K) for CePt$_3$Si decreases progressively with pressure, and its SC emerges under the Fermi-liquid state without any trace of AFM spin fluctuations. 
In order to gain insight into this contrasted SC characteristics of both compounds, we would remark a multiband effect in HF systems by systematically comparing with most of HF SC compounds reported thus far. 

A number of studies on HF compounds have revealed that unconventional SC arises at or close to the QCP, where the magnetic order disappears at low temperatures as the function of $P$. These findings suggest that the mechanism forming Cooper pairs can be magnetic in origin. However, the natures of SC and magnetism are still unclear when SC appears very close to AFM. In fact,  the $P$-induced first-order transition from AFM to paramagnetism has been revealed in CeIn$_3$, \cite{Mathur98,SKawasaki2008} CePd$_2$Si$_2$, \cite{Grosche} and CeRh$_2$Si$_2$ \cite{Movshovic96,Araki} near the boundary where SC emerges without the development of AFM spin fluctuations. 
For these compounds - denoted as group [I] compounds - that exhibit relatively high $T_N$ and low $T_c$ \cite{Kitaoka1}, a multiband model was proposed considering that the FS relevant with the AFM order differs from that with SC, \cite{SKawasaki2008} similar to the case of CePt$_3$Si.  
Remarkably, a different behavior was reported in the archetypal HF superconductors - CeCu$_2$Si$_2$ \cite{Steglich,Jaccard92,YKawasaki1,YKawasaki2} and CeRhIn$_5$\cite{Hegger00,Muramatsu1,YashimaPRB2007} - with relatively low $T_N$ and high $T_c$ \cite{Kitaoka1} (denoted as group [II] compounds). Although an analogous behavior relevant with an AFM-QCP has been demonstrated in both compounds, it is noteworthy that the SC region associated with group [II] extends to higher pressures than that in group[I], their $T_c$ value reaching its maximum away from the verge of the AFM order \cite{Muramatsu1,YashimaPRB2007,Bellarbi,Thomas}.
The SC of group [II] compounds emerges under the Fermi-liquid state dominated by strong  AFM spin fluctuations. 
From a comparison of group [II] with group [I], we propose that the SC state in group [II] may be described by a single band picture, namely, both AFM and SC take place on the same band. 
Interestingly, in high-$T_c$ copper oxides where the single band picture is applicable, the phase diagram of AFM and SC has been recently established as the function of the carrier doping level,\cite{MukudaPRL2006,MukudaCuprate} resembling that of group [II] compounds.  
Although the band structure of HF compounds is generally very complicated, there exist typical experimental results that indicate the validity of the single band picture where AFM and SC occur commonly. 
For example, (1) the dHvA experiment with CeRhIn$_5$ under $P$ \cite{Shishido} revealed that the FS topology dramatically changes at the QCP where the localized-$f$ electronic states turn into the coherent HF bands. (2) The corresponding magneto-transport measurements revealed the anomaly at the QCP on CeMIn$_5$ (M = Rh and Co)\cite{Nakajima} and also the high-$T_c$ cuprates. (3) In the uniformly coexisting phase of AFM and SC of CeRhIn$_5$ under $P=$ 1.6 GPa, the NQR measurement revealed that the extremely large RDOS remained at the Fermi level below $T_c$ because of an intimate coupling between the AFM order parameter and the SC order parameter \cite{YashimaPRB2007}, whereas the RDOS became very small in the single SC phase at $P\sim$ 2 GPa. 
In contrast, in the case of group [I], the HF bands remaining below $T_N$ are responsible for the unconventional SC with the line-node gap, as well as in group [II]; however, a very small residual DOS at $E_F$ was observed well below $T_c$, even in the uniformly coexisting phase of AFM and SC for CeIn$_3$\cite{SKawasaki2008}, UPd$_2$Al$_3$\cite{Kyogaku}, URu$_2$Si$_2$\cite{Kohori}, and CePt$_3$Si. In this context, a criterion that the single band picture is applicable would be whether a large fraction of RDOS remains well below $T_c$. 
As a result, the AFM interaction in the same band plays vital role not only for the AFM order but also for the strong coupling SC in group [II]. 
Once the AFM order collapses above QCP, the AFM order parameter will totally evolve into the SC order parameter, resulting in relatively high SC transition around QCP.  Furthermore, we note that the tetracritical point of the AFM phase, the uniformly mixed phase of AFM+SC, the SC phase, and the PM phase is present at a certain temperature and pressure without a magnetic field in CeRhIn$_5$ \cite{YashimaPRB2007} and high-$T_c$ copper oxides.\cite{MukudaCuprate}
In this context, we propose that these phenomena in group [II] and high-$T_c$ copper oxides may be phenomenologically described by the SO(5) theory that unifies the AFM and SC states by a symmetry principle and describes their rich phenomenology through a single low-energy effective model.\cite{SO5} 
Therefore, on the basis of these findings, we consider that CePt$_3$Si and CeIrSi$_3$ are classified into the groups [I] and [II], respectively. 
Consequently, these considerations will lead us to a coherent understanding for the SC around the QCP of the AFM order in strongly correlated matter.

\subsection{Possible stacking faults in the high-quality single crystals}

In the CePt$_3$Si, no evidence of parity mixing due to the lack of inversion symmetry has been obtained experimentally so far, suggesting that the theoretical models for CePt$_3$Si put forth to date \cite{Hayashi} seem to be inadequate because of the unknown complicated multiband effect. 
There still remains the underlying issue regarding why the disordered domains - amounting to 30\% of the volume fraction - are present even in the highest quality single crystal ($\sharp$4) of CePt$_3$Si. 
Unexpectedly, these domains could be distinguished neither by X-ray spectroscopy nor by the resonance line shape in NMR. 
It should be noted that a such marked sample dependence of $T_c$ has never been reported in CeIrSi$_3$\cite{Sugitani} and CeRhSi$_3$\cite{Kimura} and also in other HF superconductors. 

\begin{figure}[htbp]
\begin{center}
\includegraphics[width=0.7\linewidth]{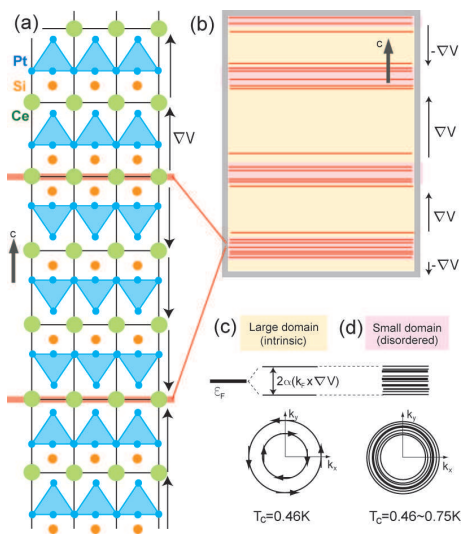}
\end{center}
\caption[]{(Color online) The present single crystal ($\sharp$4) includes a large volume fraction of homogeneous domains with $T_c=0.46$ K, including a small fraction of disordered domains with $T_c=0.8$ K. (a) As one of the possible origins of the disordered domains in this sample, we assume that stacking faults are present along the c-axis on a microscopic scale, as indicated by the red lines. Here, the inversion symmetry is seemingly recovered at an interface. (b) A large-scale view of the crystal, including dense stacking faults. (c) For the homogenous domains, the Rashba-type ASOC causes the splitting of the Fermi surfaces due to the lack of inversion symmetry. (d) The ASOC is reduced as a result of an average over interfaces for dense stacking faults with many interfaces. This may also make the disordered domains break up the coherent HF bands to suppress the unconventional SC even in the single crystal. }
\label{fig:domain}
\end{figure}

One of possible explanation for the nonnegligible amount of the disordered domain in high-quality single crystal is the presence of stacking faults that are inevitably introduced in the process of crystal growth.  As illustrated in Figs. \ref{fig:domain}(a) and \ref{fig:domain}(b), the stacking faults are assumed to be present along the c-axis on a microscopic scale; the intrinsic and disordered domains of the high-quality single crystal are derived from large twin domains with less interfaces and small twin domains with many interfaces, respectively. 
In this case, the inversion symmetry is expected to be seemingly recovered at the interfaces of the stacking fault, resulting in the reduction and/or distribution of the Rashba-type ASOC locally. 
In the homogeneous domains, an asymmetric potential gradient $\vec{E}=\nabla V $//(001) brings about the Rashba-type ASOC ${\cal H}_{SO}\sim \alpha (k\times \nabla V)\cdot \sigma$ \cite{Gorkov}. Here, $\alpha$ is the coupling constant of the ASOC. This causes the splitting of the Fermi surfaces (FSs) into two pieces; the electron spins align clockwise on one FS and anticlockwise on the other FS so as to be parallel to $k_F\times\nabla V$, where $k_F$ is the Fermi momentum, as illustrated in Fig. \ref{fig:domain}(c). In the disordered domains, however, the asymmetric potential gradient is locally reduced and/or distributed, which causes a possible variation in the amplitude and sign of $\nabla V$. As illustrated in Fig. \ref{fig:domain}(d), this event is anticipated to reduce and/or distribute the Rashba-type ASOC and hence to give a degeneracy again on the clockwise and anticlockwise spin states on these FSs.  
This may also act as one of the scattering sources that causes quasiparticles to break up the $f$-electron-derived coherent HF bands as well as the disorders induced in the well-ground sample do.

\section{Conclusion}

$^{195}$Pt-NMR studies on CePt$_3$Si have revealed the presence of homogeneous and disordered domains even in a high-quality single crystal. Homogeneous domains inherent to CePt$_3$Si exhibit unconventional SC with a line-node gap below $T_c=$ 0.46 K. In these domains, the coherent HF bands, which are formed through the hybridization between $f$-electrons and conduction electrons, are responsible for the unconventional SC. 
In contrast, the disordered domains reveal the conventional BCS $s$-wave SC with a high $T_c$ of 0.8 K, which is analogous to $T_c$= 0.6 K for LaPt$_3$Si. 
In the disordered domains, the conduction bands existing commonly in LaPt$_3$Si may be responsible for the conventional $s$-wave SC state possibly mediated by the electron-phonon interaction.  
In this context, the small peak in $1/T_1$ just below $T_c$ reported in the previous paper \cite{Yogi2004} is not due to a two-component order parameter composed of spin-singlet and spin-triplet Cooper pairing states, but due to the contamination of the disorder domains that are in the $s$-wave SC state.

A $^{29}$Si-NMR study under pressure has revealed that the collapse of the AFM order does not significantly change the fraction of the DOS at $E_{\rm F}$ that is responsible for the onset of SC. This result provides sound evidence that the Fermi surfaces that are responsible for the SC state differ from those responsible for the AFM order. On the basis of these results, we propose that these unusual SC and magnetic characteristics of CePt$_3$Si can be described by a multiband model in which it is considered that the impurity scattering is caused by the disorders and/or that the stacking faults break up the 4f-electrons-derived coherent HF bands, but not others. In order to gain further insight into the possible order parameter in CePt$_3$Si, Knight-shift measurement in the SC state using a better single crystal is highly desired.

\section*{Acknowledgement}

We would like to thank S. Fujimoto and Y. Yanase for their valuable comments. 
This work was supported by a Grant-in-Aid for Specially Promoted Research (20001004) and by the Global COE Program (Core Research and Engineering of Advanced Materials-Interdisciplinary Education Center for Materials Science) from the Ministry of Education, Culture, Sports, Science and Technology (MEXT), Japan.


\begin{table}[htbp]
\caption[]{CePt$_3$Si samples used in our NMR studies. The previous NMR studies by Yogi et al.\cite{Yogi2004,Yogi2006} were performed mainly for the single crystal $\sharp$2 \cite{Yasuda}.}
\label{samples}
\begin{center}
  \begin{fulltabular}{cllll}
    \hline
Sample No. & Crystal & $T_c^{SH}$$^\dagger$ & $T_c^{R}$$^\ddagger$ & References \\
    \hline
$\sharp$1  & poly. (coarse powder)   & $\sim0.75$ K   & $\sim 0.75$ K   &  Bauer G.\cite{Bauer}  \\
$\sharp$2  & single (coarse powder)  & $0.4\sim$0.8 K  & $\sim 0.75$ K  & \={O}nuki G.\cite{Yasuda,Yogi2004,Yogi2006}   \\
$\sharp$3a & poly.$^*$ (coarse powder)   & $0.4\sim$0.8 K   & $\sim0.75$ K  & \={O}nuki G.   \\ 
$\sharp$3b & poly.$^*$ (well-ground powder)& $\sim$0.75 K   & $\sim0.75$ K   & \={O}nuki G.   \\
$\sharp$4  & single (coarse powder)  & $\sim0.46$ K   &  $\sim0.75$ K   & \={O}nuki G.\cite{Takeuchi}   \\
    \hline
    \end{fulltabular}
 \end{center}
\footnotesize{$*$)$^{29}$Si-enriched sample.}\\ 
\footnotesize{$\dagger$) $T_c^{SH}$: estimated from a specific heat jump}\\
\footnotesize{$\ddagger$) $T_c^{R}$: estimated from  a drop in the resistivity }
\end{table}

\end{document}